# Superconductor-insulator transition in nanowires and nanowire arrays

**J E Mooij**[1,2], **G Schön**[3], **A Shnirman**[4,5], **T Fuse**[2,6], **C J P M Harmans**[2], **H Rotzinger**[1] **and A H Verbruggen**[2]

1. Physikalisches Institut, Karlsruhe Institute of Technology, Wolfgang-Gaede-Str. 1, 76131 Karlsruhe, Germany
2. Kavli Institute of Nanoscience, Delft University of Technology, 2628 CJ Delft, The Netherlands
3. Institut für Theoretische Festkörperphysik, Karlsruhe Institute of Technology, Wolfgang-Gaede-Str. 1, D-76131 Karlsruhe, Germany
4. Institut für Theorie der Kondensierten Materie, Karlsruhe Institute of Technology, Wolfgang-Gaede-Str. 1, D-76131 Karlsruhe, Germany
5. L.D. Landau Institute for Theoretical Physics RAS, Kosygina street 2, 119334 Moscow, Russia
6. Advanced ICT Research Institute, National Institute of Information and Communication Technology, 4-2-1, Nukui-Kitamachi, Koganei, Tokyo 184-8795, Japan

E-mail: j.e.mooij@tudelft.nl

**Abstract.** Superconducting nanowires are the dual elements to Josephson junctions, with quantum phase-slip processes replacing the tunneling of Cooper pairs. When the quantum phase-slip amplitude $E_S$ is much smaller than the inductive energy $E_L$, the nanowire responds as a superconducting inductor. When the inductive energy is small, the response is capacitive. The crossover at low temperatures as a function of $E_S/E_L$ is discussed and compared with earlier experimental results. For one-dimensional and two-dimensional arrays of nanowires quantum phase transitions are expected as a function of $E_S/E_L$. They can be tuned by a homogeneous magnetic frustration.

## 1. Introduction

Superconducting nanowires are much more interesting than their linear current-phase relation implies, namely because of the phenomenon of quantum phase-slip (QPS). As a result, nanowires are nonlinear elements which are dual to Josephson junctions, with the roles of phase and charge, and simultaneously current and voltage, being interchanged. In this paper we consider single wires as well as one- and two-dimensional wire arrays. We explore the consequences of QPS for the crossover or zero-temperature phase transitions from superconducting (inductive) to insulating (capacitive) behaviour when the strength of QPS is increased. The transitions can be tuned by a magnetic frustration. Charge disorder, which washes out many of the interesting properties of the phase diagram in the case of Josephson junction arrays, plays a different role. We compare with existing experiments on single wires and make predictions for the arrays that can be tested in experiments.

We consider homogeneous superconducting nanowires with small cross-section and high normal-state resistance. A current through the wire varies linearly with the gauge invariant phase difference $\varphi$ according to $I = \varphi \Phi_0 / 2\pi L$ with $\Phi_0 = h/2e$ being the flux quantum and $L$ the length-dependent kinetic inductance. In these weak wires phase-slip processes occur, as a result of which the phase difference flips by $2\pi$. The process can also be viewed as the crossing of a $2\pi$ fluxoid. An individual phase-slip takes place in a region of size roughly equal to the coherence length, in a time of the order of the inverse gap $h/2\Delta$. There is an energy barrier $E_B$ that is approximately equal to the loss of condensation energy in this region where the order parameter is temporarily suppressed. At high temperatures the barrier can be overcome by thermal activation, as has been studied extensively in theory and experiment [1]. In recent years it has become clear that at low temperatures phase-slips are possible by quantum tunneling of the fluxoid [2-5]. The quantum nature of the process implies that a superposition of the fluxoid having crossed and not having crossed the wire is conceivable. Indeed, superposition states have been predicted and observed in phase-slip flux qubits [6,7].

A superconducting nanowire with QPS can be viewed as the dual to a Josephson junction and the fluxoid as the dual particle to the Cooper pair [8]. In the Josephson junction, the tunneling Cooper pair picks up a phase factor $e^{\pm i\varphi}$ leading to a coupling energy $U = E_J(1-\cos\varphi)$. The derivative of this energy with respect to phase gives the current $I = I_0 \sin\varphi$, and the time derivative of the phase is proportional to the voltage. For the nanowire, the analogy predicts that the fluxoid picks up a factor $\exp(\pm 2\pi i Q/2e)$ when tunneling, where $Q$ is the charge that has passed through the wire. This leads to a QPS energy $U = E_S\left[1-\cos(2\pi Q/2e)\right]$ where $E_S$ is the amplitude of the QPS process. The derivative of $U$ with respect to $Q$ yields the voltage, and the time derivative of $Q$ is the current through the wire.

Phase-slips may take place all along the length of the nanowire. The whole length also determines the kinetic inductance. Both components contribute to the voltage and are effectively connected in series. Thus the nanowire is a nonlinear device, which can be represented as shown in figure 1. The voltage and current are determined by the charge $Q$ that has passed through the wire according to

$$V = V_0 \sin(2\pi Q/2e) + L\ddot{Q}, \qquad I = \dot{Q} \qquad (1)$$

The voltage scale $V_0$ follows from the QPS amplitude $E_S$ - for which an estimate will be provided below - according to

$$E_S = 2eV_0/2\pi. \qquad (2)$$

The inductance is dominated by the high kinetic one and defines the inductive energy scale

$$E_L = \Phi_0^2/2L \qquad (3)$$

The total energy of the nanowire junction then is

$$U = E_S\left[1-\cos(2\pi Q/2e)\right] + L\dot{Q}^2/2 \qquad (4)$$

Note the duality with the capacitively shunted Josephson junction with energy including the charging energy. The dual properties are listed in figure 1.

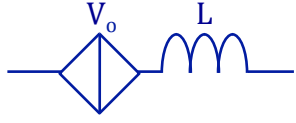 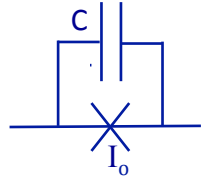

$V = V_0 \sin(2\pi Q / 2e) + L\ddot{Q}$  
$I = \dot{Q}$  
$U = E_S\{1 - \cos(2\pi Q / 2e)\} + L\dot{Q}^2 / 2$  
$V_0 = 2\pi E_S / 2e$  
$E_L = (h/2e)^2 / 2L$

$I = I_0 \sin\varphi + C(\hbar/2e)\ddot{\varphi}$  
$V = (\hbar/2e)\dot{\varphi}$  
$U = E_J(1 - \cos\varphi) + C(\hbar/2e)^2 \dot{\varphi}^2 / 2$  
$I_0 = (2e/\hbar)E_J$  
$E_C = 4e^2/2C$

**(a)** QPS junction          **(b)** Josephson junction

**Figure 1.** Circuit representation of (a) a nanowire with quantum phase-slip element in series with inductance and (b) a capacitively shunted Josephson junction. These objects are each other's dual.

As is well established, single Josephson junctions exhibit a crossover between inductive, superconducting behaviour when $E_J \gg E_C$ to capacitive, insulating behaviour when $E_C \gg E_J$. For one-dimensional serial chains or two-dimensional arrays of Josephson junctions the transition occurs as a zero-temperature quantum phase transition. The equivalent transitions are to be expected for single QPS junctions, for 1D parallel arrays and for 2D arrays of QPS junctions. Starting from the inductive, superconducting regime where $E_L \gg E_S$, increasing $E_S$ induces a transition to a capacitive insulating regime. Note that the low-$E_S$ regime of superconducting nanowires corresponds to the Coulomb blockade regime of Josephson junctions.

For Josephson junctions, an approximate duality exists between the charging and the superconducting regimes. In the charging regime for $E_C \gg E_J$, the relevant charge is induced by a gate voltage $V_g$ coupled via a gate capacitor, $Q_g = n_g 2e = C_g V_g$. The energy $E_{ch}(n_g)$ consists of a series of shifted parabolas. The Josephson coupling induces avoided crossings and leads to multiple bands $E_i(n_g)$. The energy is periodic in the induced charge, but it does not have the simple single-valued character of the potential energy $U(\varphi)$ in the inductive regime. Similarly, the QPS junction has multiple periodic bands $E_i(f)$ depending on the frustration $f$ induced by a magnetic flux through the loop. This is illustrated in figure 2.

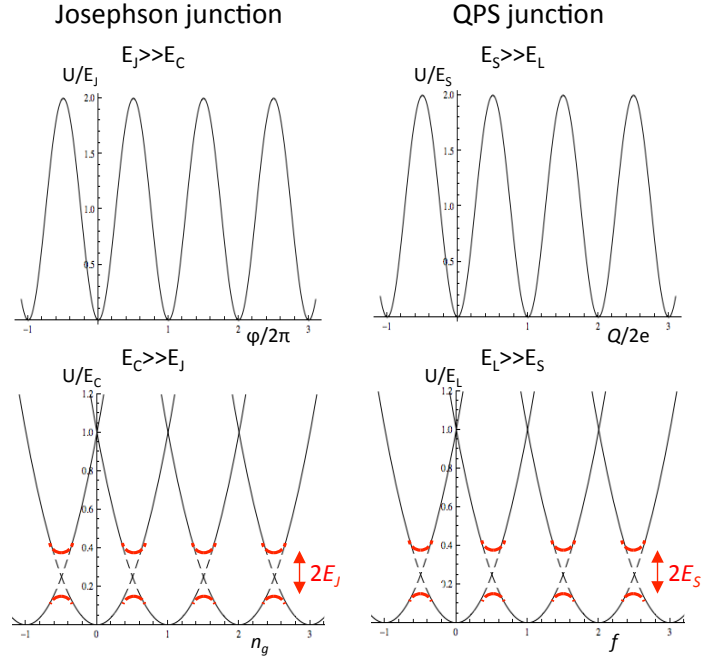

**Figure 2.** Duality of quantum phase slip junctions (QPS) and Josephson junctions. For Josephson junctions a crossover from a superconducting to an insulating behavior occurs when the charging energy $E_C$ is increased beyond $E_J$. The QPS junction has a similar crossover from insulating behavior when the QPS amplitude $E_S$ is larger than the inductive energy $E_L$, to superconducting behavior for large $E_L$. The inductive regime for QPS and the charging regime for Josephson junctions have multiple bands, that originate from shifted parabolas with avoided crossings. These parabolas are indicated with the number of fluxoids $n_f$ or the number of Cooper pairs $n_p$ that have crossed. $n_g$ is the normalized charge on a gate capacitor of a Cooper pair box, $f$ the magnetic frustration in a flux loop.

In realistic circuits, due to uncontrolled charged defects, the charges on islands between Josephson junctions have random offsets with values of the order of 2e. As a consequence, many interesting predictions made for the charge state of 1D and 2D Josephson junction arrays in the weak tunneling regime (multi-lobe structure as a function of an overall gate voltage and even super-solid phases) are washed out due to the disorder averaging. In contrast, in an array of multiple closed loops the magnetic flux is experimentally well-controlled, leading to richly structured responses as function of the applied flux. Thus the QPS junctions provide, in the weak phase-slip regime, the opportunity to study what corresponds to the weak-tunneling regime of Josephson junctions that could not be probed in experiments due to the disorder. The unavoidable charge disorder will affect the experimental study of QPS arrays only in the regime with strong phase-slip.

## 2. Estimates of $E_L$ and $E_S$

With nanowires the quantum transition between the superconducting and insulating regimes takes place for a sufficiently high value of the ratio

$$\alpha = E_S / E_L \qquad (5)$$

Quantum phase slip processes are only relevant for very weak nanowires, both in terms of the wire cross-section and the normal state resistivity of the metal. The normal state resistance per unit length is typically extremely high ($> 10\, k\Omega/\mu m$), and the mean free path of the electrons is very short. We assume that the superconducting metal follows the BCS-Gorkov theory with critical temperature $T_c$ and normal state resistance $R_n$. In this case the kinetic inductance of the wire is $L = 0.18 h R_n / k_B T_c$, [1] which leads to an inductive energy

$$E_L = \frac{\Phi_0^2}{2L} = 17.4 \frac{R_q}{R_n} k_B T_c, \tag{6}$$

with $R_q = h/4e^2 = 6.45\, k\Omega$ being the quantum resistance.

It is more difficult to find a quantitative prediction for $E_S$. At temperatures near $T_c$ thermally activated phase-slip processes are well described by the time-dependent Ginzburg-Landau equations. There is no equivalent simple set of equations for low temperatures. Arutyunov et al. [4] gave extensive discussions of the physics of the problem, but their results do not lend themselves to a direct quantitative comparison with experiment. Giordano [2] and later Lau et al. [3] used a phenomenological extrapolation of the time-dependent Ginzburg-Landau equations to zero temperature. Partly following them, we assume that individual quantum phase-slip events occur in a region of the nanowire of the size of the coherence length $\xi = (\xi_0 l)^{1/2}$ long (where $\xi_0$ is the BCS coherence length and $l$ the electronic mean free path). The tunnel barrier $E_B$ is the loss of condensation energy when the order parameter goes to zero over the volume of that region. The attempt frequency is given by the gap, $\omega_0 = \Delta/\hbar = 1.76 k_B T_c / \hbar$. Furthermore one assumes that phase-slips at one position do not influence phase-slips elsewhere and the total rate therefore simply scales with the length of the wire $A$ in units of $\xi$. The QPS amplitude then has the form $E_S = c_1 (A/\xi) \hbar \omega_0 \eta \exp(-c_2 E_B / \hbar \omega_0)$, where $c_1$ and $c_2$ are constants of order one. The factor $\eta$ is a prefactor as found in all Kramers-type calculations of quantum tunneling. Such prefactors depend on aspects such as the shape of the barrier and the damping. Lau et al. used a prefactor $(R_q/R_\xi)^{1/2}$, based on time-dependent Ginzburg-Landau calculations at higher temperatures. We follow Arutyunov et al. [4], who predict at low temperature a prefactor $\eta = R_q / R_\xi$. The dominant factor in the expression is the exponential one, which can be written as $\exp(-bR_q/R_\xi)$ using the BCS-Gorkov expressions for a dirty superconductor [3]. Here $R_\xi = \xi R_n / A$ is the resistance of the nanowire over one coherence length. Thus the result, with $a$ and $b$ being unknown constants of order one, is

$$E_S = a \frac{A}{\xi} k_B T_c \frac{R_q}{R_\xi} \exp(-b \frac{R_q}{R_\xi}) \tag{7}$$

**3. Single nanowires**

For single Josephson tunnel junctions, there is no sharp value $\alpha_c$ of the ratio $E_J/E_C$ where the crossover between the two limiting behaviours occurs, neither in theory nor in experiment. A value $0.1 \leq \alpha_c \leq 1$ is a reasonable estimate. With single QPS junctions, duality leads to an expected crossover at the same critical value of $\alpha = E_S / E_L$. When $\alpha$ is small, the wire acts as a superconductor with zero DC resistance for weak currents. When the length of the wire is increased or the cross-section reduced, the inductive energy decreases while the phase-slip rate increases. The wire will eventually become an insulator for weak DC driving. In figure 3 an example is shown of an I-V characteristic of a nanowire in the insulating regime. The sample is a NbSi wire with thickness 5 nm, width around 10 nm, length 1 µm and normal state resistance 100 kΩ. Using equation (6) the inductive energy is about $E_L/h$=26 GHz. The observed value for $V_0$ of 200 µV corresponds to $E_S/h = 16$ GHz. The ratio $\alpha = E_S/E_L$ is about 0.6, which apparently is above the critical value $\alpha_c$.

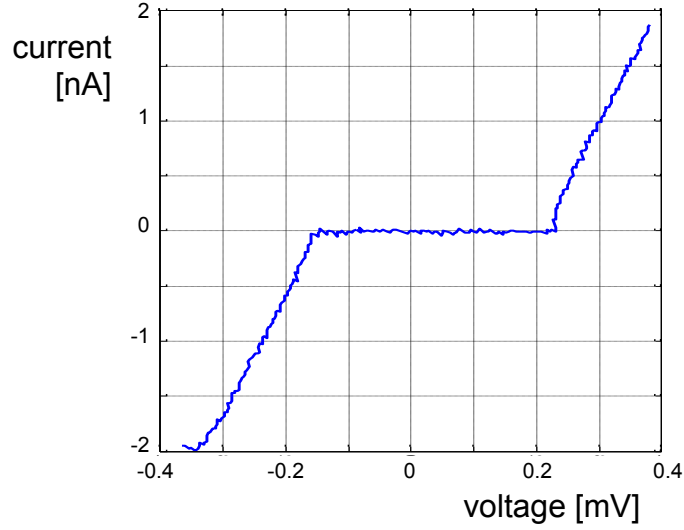

**Figure 3.** Current-voltage characteristic (voltage-biased) of a long narrow nanowire of NbSi. A critical voltage is observed of about 200 µV. A resistance of 50 kΩ was placed in series to provide a high-impedance environment; the combination of wire and resistor was voltage-biased. Due to parasitic capacitances $E_S$ may have been suppressed.

The response of the nanowire to a weak current or voltage drive will be insulating (capacitive) or superconducting (inductive), depending on the ratio of $E_S$ and $E_L$. This ratio is determined by the normalized length $\lambda \equiv A/\xi$ and resistance $r_\xi \equiv R_\xi / R_q$. With $E_S = a\lambda k_B T_c r_\xi^{-1} \exp(-b/r_\xi)$ and $E_L = 17.4 k_B T_c / (\lambda r_\xi)$ we expect the transition to occur when

$$\frac{E_S}{E_L} = \frac{a}{17.4}\lambda^2 \exp(-b/r_\xi) = \alpha_c$$

i.e., when

$$r_\xi(\lambda) = \frac{b}{\ln(a\lambda^2/17.4\alpha_c)} = \frac{b}{\ln(c\lambda^2)} \tag{8}$$

where $c = a/(17.4\alpha_c)$. For specific values of $c$ and $b$ the results for $r_\xi(\lambda)$ are shown in figure 4. We assumed $\alpha_c = 0.3$, i.e., the values $c$=0.1, 0.05, 0.025 correspond to $a$= 0.52, 0.26, 0.13.

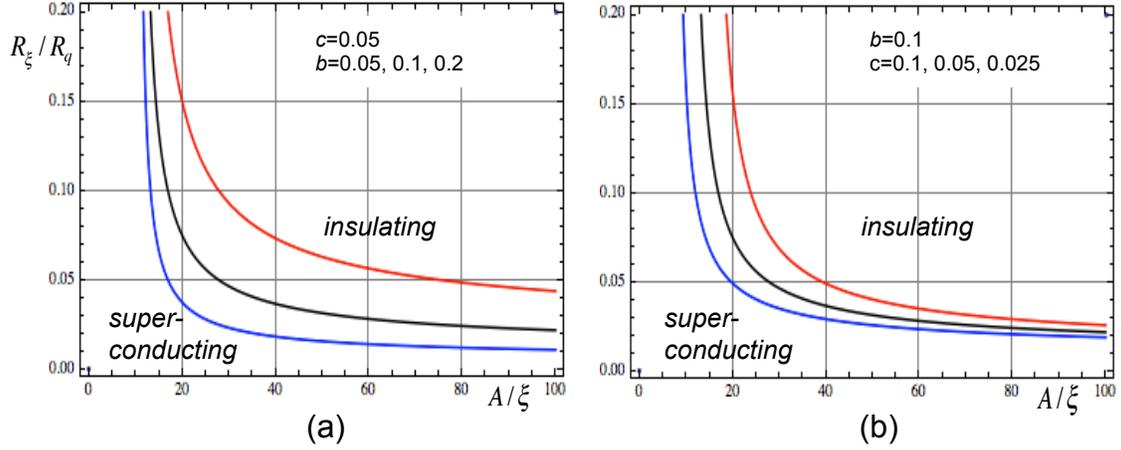

**Figure 4.** Phase boundary for a single QPS junction according to equation (8), for several values of $c$ and $b$. $A$ is the length of the wire, $\xi$ the coherence length, $R_\xi$ the wire resistance over one coherence length, and $R_q = h/4e^2$ the quantum resistance. For long wires and wires with high $R_\xi$, the response is insulating. (a) The values are $c=0.05$ and $b=0.05$ (blue), 0.1 (black), 0.2 (red). (b) The values are $b=0.1$ and $c=0.1$ (blue), 0.05 (black), and 0.025 (red).

Bollinger et al. [9] collected data for a large number of MoGe nanowires. They observed that the differential resistance at low temperatures was either very low ('superconducting') or very high ('insulating'), with no data lying in between. At the time of the publication this dichotomy could not be understood. The transition as described above provides a very natural explanation. In figure 5, the data points of Bollinger et al. have been reproduced. A value $\xi = 5$ nm was used for the conversion of wire length to $\lambda = A/\xi$ and resistance to $r_\xi = \xi R_n / A R_q$. Red squares correspond to wires with insulating response, blue triangles to superconducting wires. It is easy to find parameters for a fit to equation (8) that separates the two sets of points. The best values for this fit are $c=0.040$ and $b=0.115$. Note that for a different choice of $\xi$ an equally good fit for the phase boundary can be generated with adjusted values of $c$ and $b$ that scale with $\xi^2$ and $\xi$, respectively. The value $c=0.040$ corresponds to $a=0.21$ when $\alpha_c=0.3$ is used.

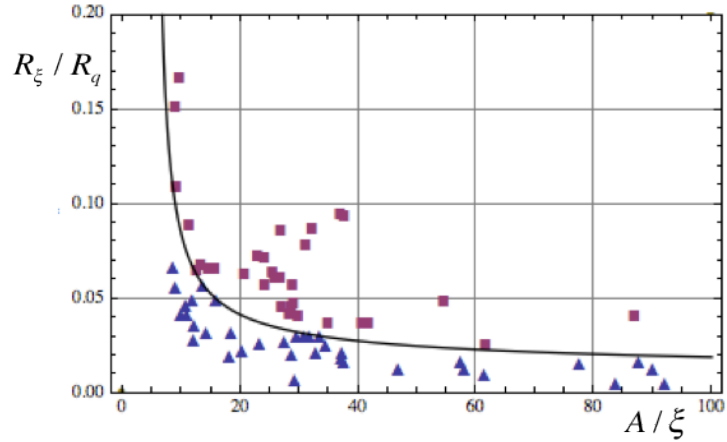

**Figure 5.** Cross-over of nanowires made from amorphous MoGe, as reported by Bollinger et al. [9]. Red squares represent wires that are insulating at low temperatures, blue triangles represent wires that in the linear response regime are superconducting. The black line follows equation (8) with $b=0.115$ and $c=0.040$.

It can be concluded that the observed transition for single nanowires from superconducting behavior to insulating behavior takes place when the ratio $E_S/E_L$ increases beyond the critical value. This crossover transition is the dual to the transition for single Josephson junctions when the ratio of Josephson energy to charging energy is varied.

**4. One-dimensional arrays of nanowires**

A one-dimensional nanowire system could be arranged either in a series or a parallel array. The former is of no particular interest; with $N$ wires in series the QPS amplitude and the inductance are both $N$ times larger, but there is no new physics. This is different for the parallel array shown in figure 6. The system has plaquettes (surrounded by a closed superconducting loop) that can contain a fluxoid; QPS allows motion of fluxoids along the length of the array. In the figure, the QPS nanowires are the vertical sections with inductance $L$ and phase-slip strength characterized by $V_0$ as before. The horizontal wires are too wide to allow phase-slip. The total 'horizontal' inductance in a cell is $L_0$. In the figure all inductances $L_0$ are pictured on the top side, but they could also be distributed between top and bottom.

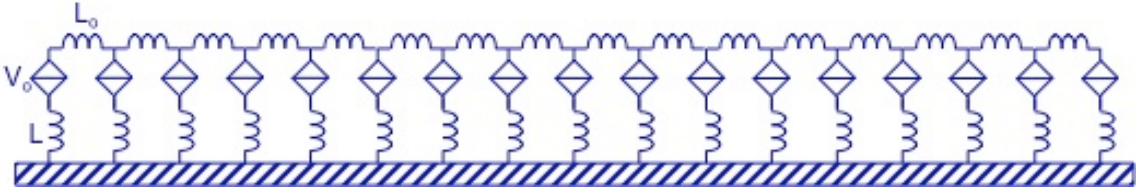

**Figure 6.** One-dimensional parallel array of nanowires. The vertical nanowires have inductance $L$ and QPS with critical voltage $V_0$. The horizontal connections allow no QPS; their inductance is $L_0$.

If $\varphi_i$ is the phase difference over vertical nanowire $i$ and $\psi_i$ is the phase difference over the adjacent horizontal connection, the sum of the gauge-invariant phase differences around plaquette $i$ of the array has to satisfy

$$-\varphi_i + \varphi_{i+1} + \psi_i = 2\pi(f_i - n_i) \tag{9}$$

where $f_i = \Phi_i/\Phi_0$ is the normalized flux and $n_i$ the fluxoid number for the plaquette. Also, the currents at each node add up to zero. From these, the phase distribution can be calculated. In a quasi-continuous approximation at zero frustration one finds that the nanowire phases satisfy the equation $d^2\varphi/dx^2 = (L_0/L)\varphi$, where the dimensionless position $x$ replaces the cell number. Near a local disturbance such as a fluxoid, the phase falls off as $\exp(-x/\lambda_s^f)$ with a screening length

$$\lambda_s^f = \sqrt{L/L_0} \tag{10}$$

In figure 7 the phase distribution is plotted for an array of 20 wires, calculated by minimizing the total inductive energy. Results are given for zero frustration and for $f = 0.05$. A fluxoid is present in the middle of the array, creating a phase jump of $2\pi$ at that position. The ratio $L_0/L$ has been varied, demonstrating that the fluxoid is a more localized object when the screening length is short. In practical fabrication it is possible to produce wire arrays with such values.

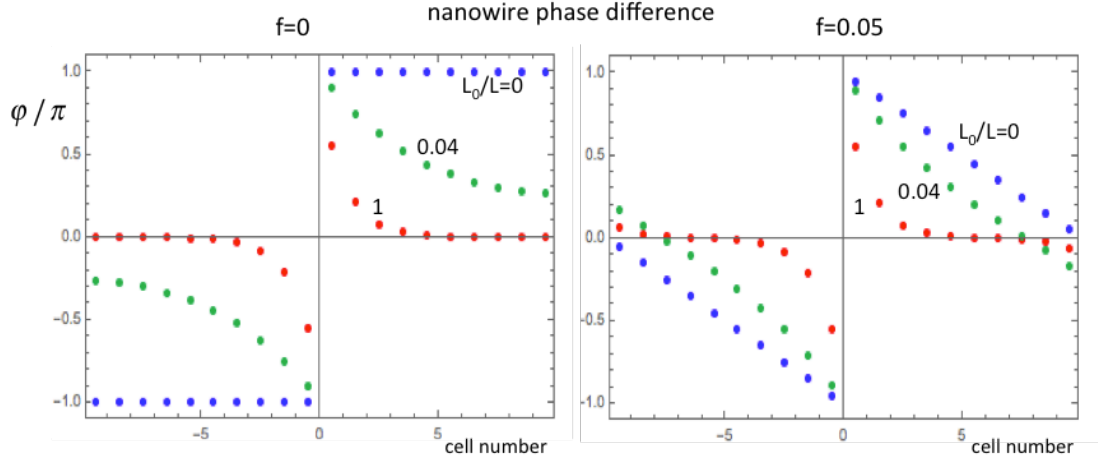

**Figure 7.** Nanowire phase differences in an array with 20 wires that contains a fluxoid in the middle, for two values of the frustration *f*.  Blue: $L_0/L$=0, green: $L_0/L$=0.04, red: $L_0/L$=1.

Quantum phase-slip processes allow fluxoids to move through the array. With an applied field, the system has its lowest energy when the fluxoid density is equal to the frustration. Fluxoids of the same (opposite) sign repel (attract) each other over a range $\lambda_s^f$. When the fluxoid density is low ($1/f \gg \lambda_s^f$), the fluxoids behave as independent particles. In the opposite limit $1/f \ll \lambda_s^f$, the fluxoids form a rigid lattice. This aspect has a strong influence on the phase transition.

The configuration of wires is the dual to a 1D chain of Josephson junctions with junction capacitance *C* and 'self-capacitance' to the ground $C_0$, where the motion of Cooper pairs is studied. The screening length for charge is $\lambda_s^c = \sqrt{C/C_0}$. The phase transition of the Josephson junction chain has been studied extensively in theory as well as in experiments [10-16]. However, there seems to be a mismatch between both when it comes to a comparison. In the one-dimensional systems most theoretical approaches concentrate on the situation when the screening length is small, which requires the islands in the array to have a self-capacitance that is larger than the junction capacitance. In practical samples the opposite is true, typically *C* is more than 100 times larger than $C_0$. Moreover, the unavoidable presence of random offset charges on the islands introduces a strong randomizing factor in the regime where the charging energy dominates. In the following we will summarize known results for Josephson junction chains and then discuss the consequences for the nanowire arrays. This will first be done assuming there is no charge or flux frustration.

Bradley and Doniach [10] analyzed quantum fluctuations for Cooper pairs in Josephson chains and concluded that with only self-capacitance $C_0$, there should at zero temperature be a quantum Berezinskii-Kosterlitz-Thouless transition involving charge-anticharge pairs. Choi, Yi, Choi, Choi and Lee [14] later gave a more extended treatment that we follow here in part. The schematic phase diagram that these authors produced is represented in figure 8; part (a) being the Josephson-chain version. The Josephson energy of the junctions is $E_J$, the charging energy of the junctions is $E_C = 4e^2/2C$, and the charging energy for the self-capacitance of the islands is $E_{C0} = 4e^2/2C_0$. The Bradley-Doniach transition takes place along the horizontal axis where $E_J/E_C=0$, at the value $E_J/E_{C0}=1.23$. Choi et al. extended the phase boundary to finite values of $C$. However, their calculations start from the low $C/C_0$ regime and may not be accurate in the experimentally important limit of large $C$ but very small $C_0$. For small values of $C_0$ where the whole chain is shorter than the screening length, the chain should behave as a single junction. One expects a crossover near $E_J/E_C=1$, as indicated by the dashed area in the figure. Chow, Delsing and Haviland [15] suggested that for high values of the screening length $\lambda_s^c$, the effective charging energy is reduced since the charge is spread over the screening length. This results in a phase boundary at $(E_J/E_C)(E_J/E_{C0}) = (2/\pi)^4$, which is indicated in figure 8 with a dashed blue line.

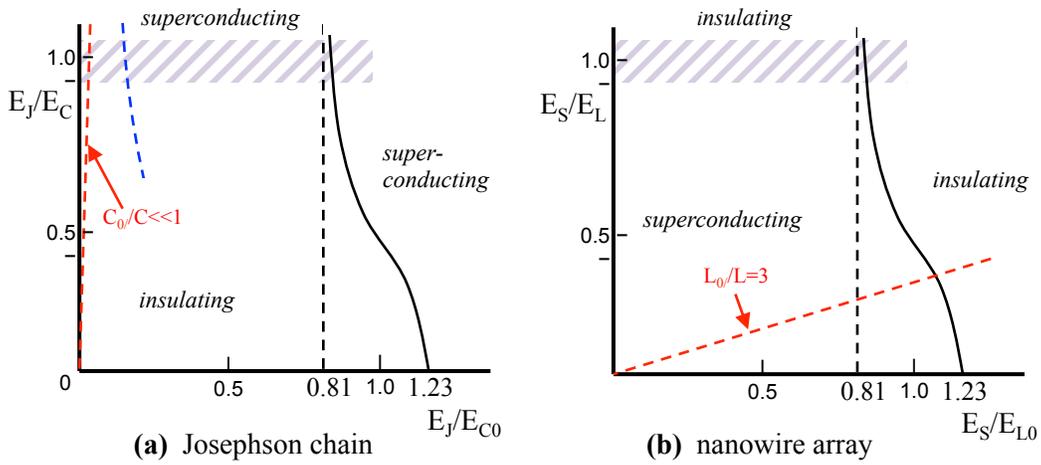

**Figure 8.** Schematic phase diagrams. (a) Josephson junction chain, $E_C = 4e^2/2C$ and $E_{C0} = 4e^2/2C_0$. Bradley-Doniach predicted the transition along the horizontal axis where $C_0/C \gg 1$, but in real samples $C_0/C \ll 1$ (dashed red line). The black curved line gives the transition as predicted by Choi et al. [14]. The blue dashed line is the approximate transition line according to reference [15]. (b) Nanowire array, $E_L = \Phi_0^2/2L$, $E_{L0} = \Phi_0^2/2L_0$. The red dashed line is for $L_0=3L$, which can be fabricated in practice. High in both plots a hashed area indicates the crossover that is likely provided by finite size and disorder.

Haviland and co-workers [15,16] have observed a transition from insulating to metallic behaviour. In their samples, they replaced the junctions by SQUIDs, so that by applying a magnetic flux they could reduce the Josephson energy without changing the charging energies. Given the low value of $C_0/C$ in the samples, their observations are likely strongly influenced by the charge disorder and represent a crossover due to the finite sample length. Yet, the inverse dependence of the resistance on the length of the sample indicated features of the superconducting Berezinskii-Kosterlitz-Thouless phase transition.

The results for the junction arrays can be translated to the nanowire system at zero frustration. The transposed phase diagram is shown in figure 8 (b). For high values of $E_S$ the arrays become insulating. Interestingly, the inductance $L_0$ can now be made larger than the nanowire inductance $L$, resulting in a screening length smaller than 1 (as shown in the figure). One expects that a quantum phase transition occurs slightly below $E_S/E_{L0}$=1.23. Given that in the insulating state charge disorder is hard to avoid, the transition is best approached from the superconducting side.

We now turn to the frustrated arrays, first considering the Josephson junction chains with charge frustration applied to the islands between the junctions. Several theoretical papers [14,15] predicted a highly structured phase diagram as a function of the charge frustration, with different insulating phases in a multi-lobe structure characterized by various charge ordered states. However, in actual samples, each island has a random offset charge of order 1 on the scale of 2$e$. When a common gate is applied to all islands, the induced charge on individual islands will oscillate with increasing gate voltage, but the randomness remains. As a consequence, experiments could not confirm or test the predicted results.

It turns out that the nanowire array may provide the experimental test ground for the above-mentioned theories. In a one-dimensional nanowire array the phase transition is influenced by an applied magnetic field in the same way as the junction array is influenced by the gate voltage. However, it is very well possible to apply a uniform flux, so that the magnetic frustration is uniform. The frustration induced by the magnetic flux has identical influence as the gate voltage would do in a defect-free Josephson junction chain. The phase diagram, when transposed to the nanowire system, looks schematically as depicted in figure 9. Since the screening currents are small the diagram should repeat periodically with period 1 as $f$ is increased or decreased. Also, the symmetry around any half-integer value of $f$ follows directly from the combination of inversion symmetry and periodicity. The diagram exhibits lobes around $f$=0 and other integer values, with localized fluxoids and hence superconducting response. Around $f$=1/2 and other half-integers, a "Neel lobe" occurs that is based on a pattern where the currents in consecutive plaquettes alternate their directions. When $E_S/E_L$ is high, the fluxoids fuse into a superfluid and the system's response is insulating. Glazman and Larkin [13] found for the junction chain an extra lobe to the right of the Neel lobe that has a 1D Luttinger character. Nanowire arrays, which do not suffer from the strong disorder of Josephson chains, can provide the opportunity to test this and further theoretical predictions (such as, e.g., supersolids).

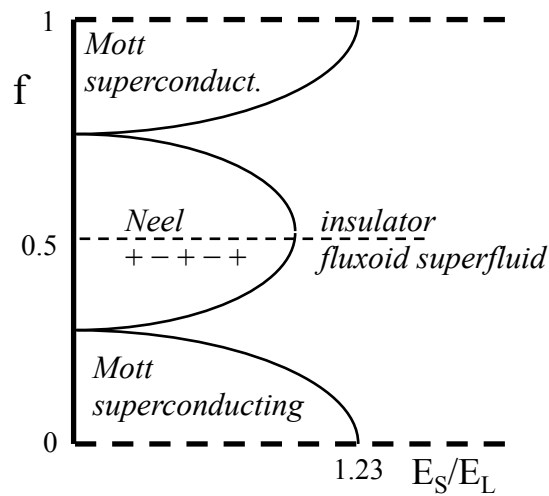

**Figure 9.** Schematic phase diagram of the frustrated one-dimensional nanowire array, transposed from the phase diagram for a Josephson junction chain according to Bruder et al. [16]. The assumption has been made that $L_0 / L \geq 1$. The diagram is periodic with period 1 and symmetric around integer and half-integer values of $f$. Lobes with a superconducting phase are positioned around integer values of $f$, the fluxoids are Mott-localized. Around half-integer values one finds a superconducting Neel phase based on alternating current directions. For high values of $E_S/E_L$ the system enters the superfluid fluxoid regime with insulating response.

**5. Two-dimensional arrays of nanowires**

Two-dimensional arrays of nanowires can be fabricated with thin-film techniques, or may be naturally present in layered materials (figure 10). The behaviour is expected to be dual to that of 2-D arrays of Josephson junctions. The array has loops around plaquettes that can be biased with magnetic flux, and which can contain fluxoids. An applied perpendicular magnetic field leads to a homogeneous magnetic frustration $f = \Phi/\Phi_o$ where $\Phi$ is the flux per plaquette. This, again, is the case because the very high inductance of the nanowires leads to a very long screening length. One could also imagine induced charge frustration by a common gate that couples to all nodes of the array. However, in practice the offset charges will wash out effects of that type of gate.

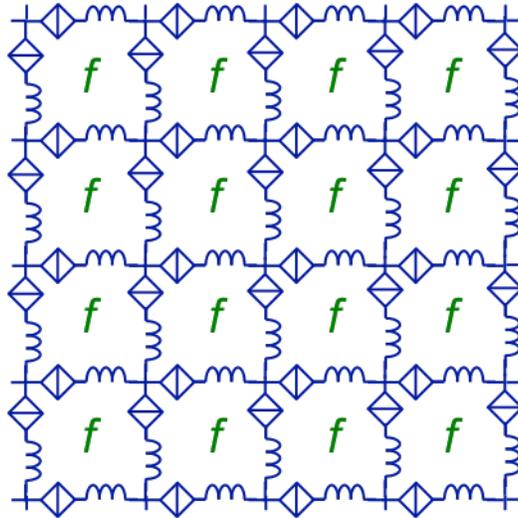

**Figure 10.** Two-dimensional nanowire array. The wires form plaquettes through which magnetic flux can pass. The frustration $f$ is the flux normalized to the flux quantum. This magnetic frustration can be made homogeneous.

For the phase differences around plaquette $i$ equation (9) still holds as in the one-dimensional array, the sum of the gauge-invariant phase differences over the wires around plaquette $i$ being equal to $2\pi(f - n_i)$ and $n_i$ is the number of fluxoids. The wires and the nodes where four wires come together have negligible capacitance. Some local charge may be induced by defects on the surface, but the value is random in fabricated arrays.

For the charge that has passed through the four wires connected to any particular node an equation similar to equation (9) is valid:

$$\sum_{j \in node} Q_j = q_{node} \qquad (12)$$

$q_{node}$ is the charge on the node and is determined by induced gate and defect charges. In practice a fabricated array has so many charged defects on its surfaces that these induced charges are fully random. No uniform charge frustration can be applied. In well-ordered layered materials the charge frustration may be relevant, but here we ignore such special effects. However, the random charge frustration may lead to an effective overall weakening of the QPS strength (not all $Q_j$ can be zero). The time derivative of equation (12) implies that the sum of all currents into a node is zero, as expected.

The two-dimensional array of nanowires as shown in figure 11 is the dual to the extensively studied array of Josephson junctions. A review of that Josephson junction work can be found in the article of Fazio and Van der Zant [16]. The Hamiltonian for the Josephson array is

$$H = \frac{1}{2}\sum_{i,i'} 4e^2 (n_i + n_i^d - n_g) C_{ii'}^{-1} (n_{i'} + n_{i'}^d - n_g) - E_J \sum_j \cos \varphi_j \qquad (13)$$

The indices $i$ or $i'$ count the islands of the array, $2en_i^d$ is the random offset charge, and $2en_g$ the imposed gate charge. $C_{ii'}$ is the capacitance matrix. The index $j$ counts the junctions. The sum of the $\gamma_j$ terms around a plaquette should confirm with the imposed flux according to equation (9). The corresponding Hamiltonian for the nanowire array is

$$H = E_L \sum_j \left(\varphi_j / 2\pi\right)^2 - E_S \sum_j \cos(Q_j / 2e) \qquad (14)$$

where $Q_j$ are the charges that have been transferred through nanowire $j$. They are subject to the restrictions of equation (12), while the $\varphi_j$ terms must obey equation (9).

The 2D Josephson array exhibits a clear zero-temperature quantum phase transition from an insulating to a superconducting state as the ratio $E_J/E_C$ is increased. Both theory and experiment were well developed, as reviewed in reference [17]. At zero frustration the transition is observed in experiment at $E_J/E_C=0.147$. Theory predicts that with increasing $E_J/E_C$, coming from the charge-ordered insulating regime there occurs a Berezinskii-Kosterlitz-Thouless (BKT) transition at $E_J / E_C = 1/(2\pi^2) = 0.051$. Coming from the phase-ordered superconducting regime a similar BKT transition should happen at the same value. In most of the theory, charge disorder is ignored, although it will strongly influence charge-anticharge pairing. Magnetic frustration yields rich structure in the phase diagram. As the fluxoid density increases with increasing $f$, one sees in transport initially an increasing effective resistance. The fluxoids or vortices move as quantum particles. At fractional values $f = m/n$ clear dips in the mobility are seen as the fluxoids are trapped by the commensurate lattice. At $f = 1/2$, this effect is so strong that properties are similar to the state around $f = 0$. With varying the charge frustration no significant effects are seen, as explained by the presence of strong charge disorder.

For nanowire arrays, one expects very similar behaviour as for the junction arrays. No experiments in the quantum phase slip regime have been performed yet, but measurements at higher temperatures with stronger wires yield results that very much resemble the data on Josephson arrays [18,19]. From the duality, the quantum phase transition at zero temperature should be expected at $E_S/E_L=0.051$.

With the Josephson arrays that have a phase-ordered regime and a charge-ordered regime on the one hand and the nanowire arrays with their phase-ordered and charge-ordered regimes on the other, a remarkable situation occurs. Ignoring charge disorder, for Josephson arrays the physics for phase excitations when $E_J$ is high and the physics for charge excitations when $E_C$ dominates are very similar. Although the corresponding terms in the Hamiltonian are not identical, one expects a high degree of duality between the two types of Josephson junction arrays. This is in particular true for small driving as is used to explore phase transitions. Similarly, the two types of nanowire arrays with either $E_S$ or $E_L$ dominating are each other's dual system. Yet, the nanowire systems are the more exact duals to the corresponding Josephson systems. Clearly, there is every reason to expect that the nanowire 2D array with $E_S/E_L \gg 1$ behaves as the Josephson junction array with $E_J/E_C \ll 1$. For both these systems, charge disorder spoils the picture and makes experiments impossible. Also, we expect that the nanowire array with $E_L/E_S \gg 1$ is almost the same as the Josephson array with $E_J/E_C \gg 1$. Charge disorder is not so important here and one studies the motion of fluxoids as driven by currents. The subtle difference between the exact cosine potential as a function of phase for the Josephson junction and the only approximate cosine-like potential for nanowires could be looked for in the experiments.

**6. Conclusions**
Nanowires and nanowire arrays that exhibit quantum phase-slip can be fabricated and studied. So far, few experimental results are available for arrays. The expectations are discussed in this paper.

For single nanowires, a crossover from an inductive superconductor to a capacitive insulator is expected and has been observed when the ratio of the phase-slip amplitude $E_S$ to the inductive energy $E_L$ is increased. The data on many nanowires that are reported on in reference [9] are consistent with the expected dependence of $E_S$ on length and resistance per unit length. The current-voltage characteristic of a nanowire in the high $E_S/E_L$ regime is the dual to the I-V characteristic of a classical Josephson junction.

One-dimensional arrays of parallel nanowires are the dual system to chains with Josephson junctions in series. Josephson chains exhibit a transition from superconducting to insulating behavior when $E_J$ is reduced, but this transition appears to be dominated by finite size effects and charge disorder. No theory is available for a phase transition in the parameter regime of the actual Josephson samples. In contrast, one can design and fabricate 1-D nanowire arrays with a screening length that is smaller than one element and one should be able to access the quantum phase transition as predicted for Josephson chains in the limit where the self-capacitance dominates over that between neighboring islands. Frustration for Josephson chains can be tuned by a gate voltage; but given the strong charge disorder this is completely ineffective. For nanowire arrays, frustration comes from a magnetic flux that is homogeneous over the array. The combination of short screening length and disorder-free frustration opens up the possibility to study in experiment what the extensive theoretical literature on Josephson junction arrays has provided in the past.

Two-dimensional arrays of nanowires with quantum phase-slip will exhibit a quantum phase transition as a function of $E_S/E_L$. This transition will be closely related to the phase transition as predicted and observed for two-dimensional Josephson arrays. The regime with strong phase-slip is formally the dual of the Josephson array with strong Josephson coupling. Arrays in that regime will exhibit insulating behaviour that is similar to the response of Josephson arrays with high charging energy. In fabricated arrays, charge disorder will inhibit detailed analysis. Nanowire arrays with weak phase-slip are the dual of Josephson arrays in the charging regime. They are expected to behave very similarly to Josephson junction arrays with strong Josephson coupling. Quantum phase-slip allows the motion of fluxoids (vortices); these objects have a logarithmic interaction potential. With magnetic frustration the density of fluxoids can be controlled; for specific values of the frustration the lattice of interacting vortices is commensurate with the nanowire lattice and the mobility is decreased. In crystals of layered materials, two-dimensional (sub)nanowire arrays with quantum phase-slip may occur without charge disorder.

**Acknowledgements**

This work was supported by the DFG Center for Functional Nano-structures (CFN) Karlsruhe and the DFG Research Unit 960 Quantum Phase Transitions, as well as by the European Union in the project SQUBIT. The theory analysis of 1-D arrays was funded by the Russian Science Foundation under grant No. 14-42-00044. We profited from discussions with many colleagues.